\documentclass[lettersize,journal]{IEEEtran}
\usepackage{amsmath,amsfonts}
\usepackage{algorithmic}
\usepackage{algorithm}
\usepackage{array}
\usepackage[caption=false,font=normalsize,labelfont=sf,textfont=sf]{subfig}
\usepackage{textcomp}
\usepackage{stfloats}
\usepackage{url}
\usepackage{verbatim}
\usepackage{graphicx}
\usepackage{cite}
\usepackage{adjustbox}

\graphicspath{{figures/}} 

\begin{document}

\title{Programmable Photonic Circuit for Optical Logic Operations and 2-Bit Decoding}

\author{Noel Francisco Prado Bucaro, Yushan Hu, Satoshi Sunada, Takeo Maruyama

\thanks{This study was supported by JSPS KAKENHI (grant number: JP22H05198).
All authors are with Kanazawa University, Kanazawa, Japan, e-mail:
(noelfpb@stu.kanazawa-u.ac.jp).}}


\maketitle

\begin{abstract}
We present a programmable silicon photonic circuit composed of cascaded multiport directional couplers interleaved with thermo-optic phase shifters. The device forms a reconfigurable interferometric network capable of realizing arbitrary $N \times N$ unitary transformations. By tuning the phase shifters, the same circuit can be configured to perform both optical logic gates and 2-bit decoding functions. Optimal phase settings are determined through Bayesian optimization. The circuit maintains stable operation across multiple wavelengths with 50 GHz spacing, suggesting potential for wavelength-parallel optical processing. These results highlight a pathway toward programmable and integrated photonic circuits.
\end{abstract}


\section{Introduction}

Optical computing has gained growing attention as electronic logic circuits approach physical limits in speed and energy efficiency. Interconnect delays and heat dissipation increasingly restrict scaling in electronics, while optical systems eliminate resistive-capacitive bottlenecks and enable ultrafast signal propagation~\cite{Miller2010,John2010}. By harnessing properties such as phase, polarization, amplitude, and wavelength, optical logic promises higher bandwidth and lower power operation than conventional electronics~\cite{oes-2022-0010-ShaoLiyang}. These characteristics make optics an appealing candidate for future computing and signal processing architectures~\cite{Singh2020}.

Within this context, digital logic is especially important. Optical implementations of logic gates provide the building blocks for more complex circuits and all-optical processors. A wide variety of approaches have been proposed. Photonic crystal waveguides achieve compact devices using interference~\cite{Caballero:22}, semiconductor optical amplifiers support high-speed operations~\cite{Chen:16,Hamie2002,Dong2009,Chan2004}, and microring resonators offer low power consumption in small footprints~\cite{Xu:07}. Plasmonic and metasurface devices push switching into the nanoscale regime~\cite{Wei2011,XuJ2024,Wang:23,Fu2012}, and diffractive optical neural networks demonstrate parallel reconfigurable operations using trained layers~\cite{Zarei2022,Ding://doi.org/10.1002/adma.202308993}. Additional progress has come from interferometric circuits on silicon~\cite{Aikawa2018,Aikawa2022} and hybrid material platforms with strong optical nonlinearities~\cite{Wu2022,He2024}. Despite these advances, most gate demonstrations struggle with trade-offs in footprint, power, and reconfigurability~\cite{Peng2018,Kita2020,He2022,Zhang2023,Anagha2022}.

Beyond gates, the decoder is another fundamental digital function. A 2-bit decoder transforms binary inputs into distinct outputs, enabling addressing, and control in larger circuits. Optical decoders have been realized using passive structures~\cite{Aikawa:24} and multimode interferometers~\cite{mmiDecoder}. While effective, these approaches are fixed by geometry and cannot be dynamically reprogrammed, which limits their integration into flexible photonic processors.

This work addresses these limitations by introducing a reconfigurable platform for both logic gates and a 2-bit decoder. Our architecture is based on cascaded multiport directional couplers interleaved with thermo-optic phase shifters. Unlike conventional $2\times 2$ couplers, multiport devices perform unitary transformations across several channels simultaneously~\cite{8846043,TangTenPort}, and the tunable phase arrays provide reconfigurable control throughout the network. As a result, a single photonic circuit can switch between logic and decoding operations on demand.

To reliably configure the device, we employ a closed-loop Bayesian optimization algorithm that tunes phase shifter voltages to maximize extinction ratios and output fidelity. This approach compensates for fabrication imperfections and wavelength variations, improving robustness in practical operation. The proposed system contributes toward scalable integrated photonic circuits for optical computing~\cite{Bogaerts2020}.

\section{Operating Principles}

A multiport directional coupler (DC) enables optical power exchange among several parallel waveguides through evanescent field interaction. In contrast to conventional $2\times2$ DCs, an $N$-port coupler consists of $N$ parallel straight waveguides with uniform spacing $G$ and coupling length $L$~\cite{TangTenPort}. This configuration allows simultaneous interactions across all channels, which can be described by an $N\times N$ complex transfer matrix $\mathbf{D}$.

When cascaded with tunable phase shifter arrays, the overall optical transformation of the system is expressed as
\begin{equation}
\mathbf{T} = \boldsymbol{\Phi}_M \mathbf{D} \cdot \boldsymbol{\Phi}_{M-1} \mathbf{D} \cdots \boldsymbol{\Phi}_1 \mathbf{D},
\label{eq:total_transformation}
\end{equation}
where $\boldsymbol{\Phi}_i$ denotes the diagonal matrix of the $i$-th phase shifter array and $M$ is the number of stages~\cite{8846043}. 

Each phase shifter array introduces independent phase delays to the $N$ channels, with diagonal elements of the form $e^{j\phi_n}$. Because these matrices are unitary, they preserve the overall unitarity of the cascaded transformation~\cite{Miller:13}. By jointly tuning the phases across multiple stages, the system can approximate arbitrary $N\times N$ unitary transformations~\cite{Clements:16,Bogaerts2020}.
While the optical propagation through the mesh is inherently linear, reconfigurable logic and decoder operations are enabled by mapping input bit patterns to specific constructive or destructive interference states. These optical states are subsequently converted into digital logic levels through the nonlinear square-law response of the photodetectors and the application of a decision threshold. 

The quality of the programmed transformation is assessed using the extinction ratio (ER) metric adapted from~\cite{mmiDecoder}, suitable for multi-output devices. For an input bit pattern $(A,B)$ producing an output power distribution $O^{AB}$, the extinction ratio between the dominant port $i$ (the intended “on” output) and the next largest port $j$ is defined as
\begin{equation}
ER^{AB} = 10 \log_{10} \left( \frac{O_i^{AB}}{O_j^{AB}} \right).
\end{equation}
A higher extinction ratio indicates stronger output contrast and more reliable logic or decoding performance.

\section{Device Fabrication and Experimental Results}

\subsection{Device Fabrication}

\begin{figure}[h!]
    \centering
    \includegraphics[width=1\columnwidth]{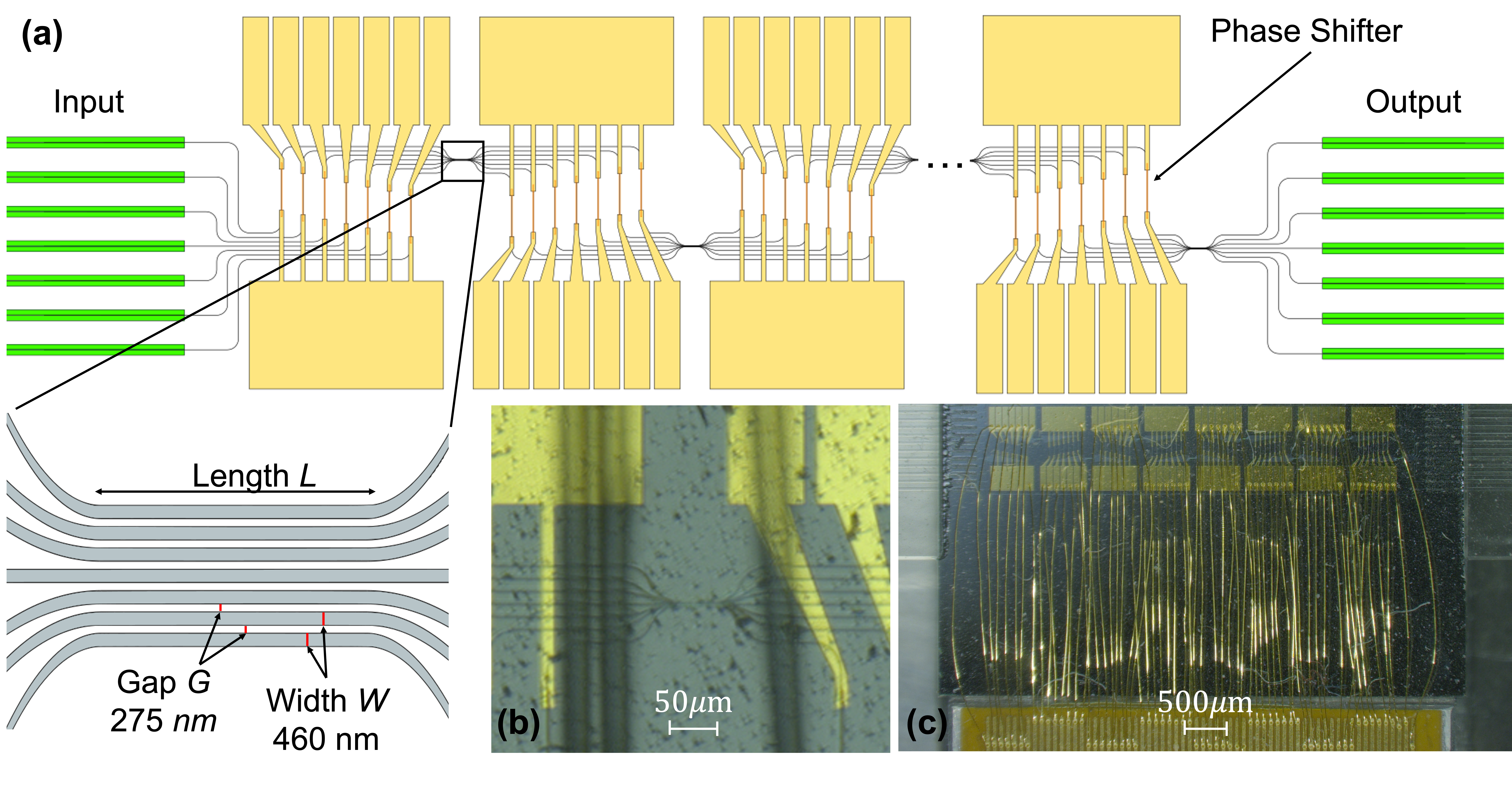}
    \caption{Design and Implementation of the Programmable Photonic Circuit. (a) Schematic architecture of the cascaded photonic mesh. The circuit consists of repeating stages of multiport directional couplers (MDCs) interleaved by thermo-optic phase shifters. The lower inset details the directional coupler geometry, defined by the interaction length (L), waveguide gap (G), and width (W). (b) Micrograph of the fabricated device region corresponding to the highlighted box in (a). The thermo-optic phase shifter pads (yellow) are positioned over the silicon waveguides to enable localized phase tuning. (c) Photograph of the fully packaged system. The photonic integrated circuit (PIC) is wire-bonded to a custom printed circuit board (PCB) to provide electronic control signals to the 56 integrated heaters.}
    \label{fig:cad_view}
\end{figure}

The proposed device consists of cascaded multiport directional couplers fabricated on a silicon photonic device. Each coupler stage comprises 7 parallel waveguides with spacing $G = 275\,\text{nm}$, width $W = 460\,\text{nm}$, and coupling length $L = 50\,\mu\text{m}$, as illustrated in Fig.~\ref{fig:cad_view}. The complete circuit includes 8 cascaded stages, resulting in a 7×7 port architecture.

A total of 56 thermo-optic heaters are integrated across the device, of which 37 are actively used in these experiments: 35 for optimization and 2 for input signal encoding. This subset was selected to provide sufficient degrees of freedom for the transformation, while maintaining a manageable parameter space for the optimization algorithm. All heaters are independently addressable with control voltages ranging from 0.1\,V to 4.9\,V.

\subsection{Device Characterization}
To evaluate the optical quality of the programmable mesh, we measured the insertion loss at 1550\,nm. The measurement was normalized to a straight reference waveguide fabricated on the same device to exclude fiber-to-chip coupling losses, isolating the performance of the photonic circuit itself. The total on-chip excess loss was measured to be 3.6\,dB.

The thermo-optic heaters were driven with voltages up to 4.9\,V. With an average heater resistance of $\approx 300\,\Omega$, the maximum power consumption per heater is approximately 80\,mW. 

Due to the coupled nature of the multiport interferometer mesh, the specific power for a $\pi$ phase shift is challenging to determine; however, experimental observations of output fringe visibility indicate that it typically falls within the 30--50\,mW range. 

Regarding thermal crosstalk, while the waveguide spacing allows for thermal interaction between adjacent phase shifters, the closed-loop optimization method inherently compensates for these effects by optimizing the global output response rather than individual heater phases.

\begin{figure}[h!]
    \centering
    \includegraphics[width=0.95\columnwidth]{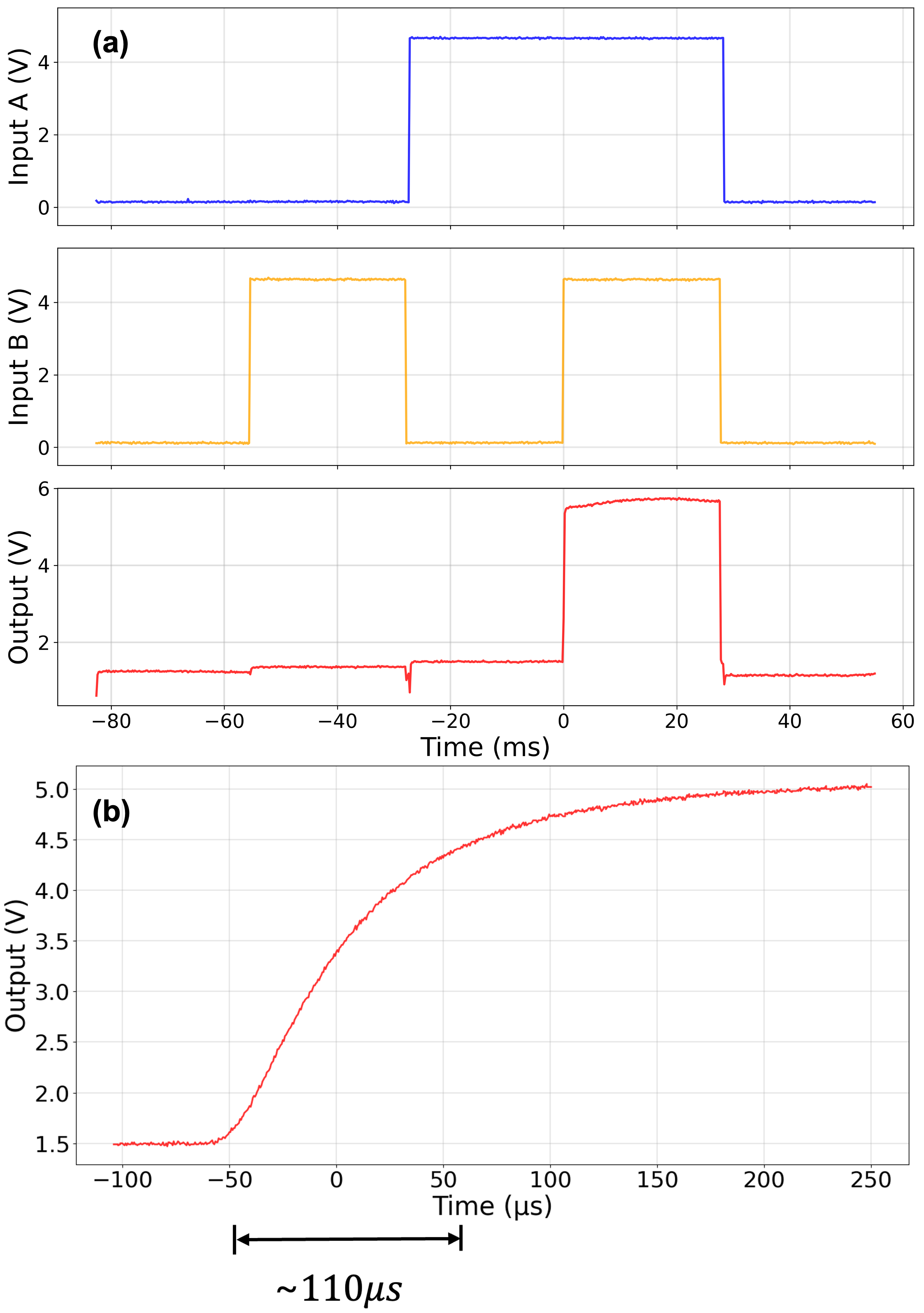}
    \caption{Dynamic response of the optical logic gate. (a) Time-domain traces showing the output following the input logic states (AND gate configuration). (b) Detailed view of the switching transient, showing a 10--90\% rise time of approximately 110\,$\mu$s, characteristic of the thermo-optic effect.}
    \label{fig:dynamic_response}
\end{figure}

We further characterized the dynamic response of the device. Since the current implementation relies on thermo-optic phase shifters for both circuit configuration and input signal encoding, the switching speed is governed by the thermal time constant of the heaters. Figure~\ref{fig:dynamic_response} shows the transient response of the logic gate output. The measured 10--90\% rise time is approximately 110\,$\mu$s. While this speed is sufficient for reconfigurable circuit programming, it limits the data rate in the current setup. However, the underlying multiport directional coupler architecture is compatible with high-speed modulation. Future iterations could replace the thermal tuners with electro-optic phase shifters based on carrier injection, which would enable operation speeds in the GHz regime while maintaining the same programmable logic functionality \cite{Reed2010}.

 A summary of the device characteristics is presented in Table~\ref{tab:device_char}.

 \begin{table}[h!]
    \caption{Device Characterization Summary}
    \label{tab:device_char}
    \centering
    \begin{tabular}{lc}
        \hline
        \textbf{Parameter} & \textbf{Value} \\
        \hline
        Operating Wavelength & 1550\,nm \\
        On-Chip Insertion Loss & 3.6\,dB \\
        Max. Power per Heater & $\sim$80\,mW \\
        Switching Speed (10--90\%) & 110\,$\mu$s \\
        \hline
    \end{tabular}
\end{table}

\subsection{Optimization Method}

To navigate the high-dimensional parameter space of the 37 active phase shifters, we employed a closed-loop Bayesian optimization, which is particularly effective for tuning complex models where evaluations are expensive \cite{snoek:1}. The optimization task is formulated as finding a voltage vector $\mathbf{v}$ that maximizes a multi-objective Figure of Merit (FoM), $J(\mathbf{v})$, designed to prioritize reliable digital logic operation. 

The objective function $J(\mathbf{v})$ is defined as the sum of weighted scores:
\begin{equation}
J(\mathbf{v}) = S_{er} + S_{str} + S_{cons}
\label{eq:fom_logic}
\end{equation}

The components of the Figure of Merit are defined as follows. First, $S_{er}$ represents the worst-case extinction ratio between the minimum ``high'' output and maximum ``low'' output. To ensure an efficient search, the $S_{er}$ mapping is intentionally non-linear: it utilizes a high-gradient region to accelerate convergence once a marginal logic gap is established, followed by an asymptotic saturation at high extinction ratios. This saturation prevents the optimizer from over-tuning a single logic state at the expense of other parameters. Second, $S_{str}$ (Signal Strength) rewards high absolute optical power to guarantee a sufficient signal-to-noise ratio. Third, $S_{cons}$ (Group Consistency) ensures that all logical output states are assigned similar signal levels to maintain predictable operation. 

The surrogate model for the photonic circuit is a Gaussian Process regressor utilizing a Matérn kernel with a smoothness parameter of $\nu=1.5$. This kernel was selected because it represents a once-differentiable process, which is better suited for modeling physical hardware responses \cite{Rasmussen:06}.

The optimization follows a multi-phase routine to balance global exploration with local refinement. Initially, the objective landscape is mapped using Latin Hypercube Sampling (LHS) supplemented by heuristic seed patterns, such as binary extremes and voltage gradients. In the subsequent global optimization phase, the Gaussian Process model evaluates 1,500 candidate configurations per cycle using an adaptive Upper Confidence Bound (UCB) acquisition function; here, the exploration parameter $\beta$ is dynamically reduced as the FoM improves to transition the search from exploration to exploitation. Finally, to prevent convergence to local optima and fine-tune the interference states, the algorithm performs stochastic local exploration by applying minor perturbations to the best-known heater configurations. This hybrid approach minimized the total number of hardware evaluations while successfully finding stable configurations within 5--10 minutes.

\subsection{Experimental Setup}

\begin{figure}[h!]
    \centering
    \includegraphics[width=1\columnwidth]{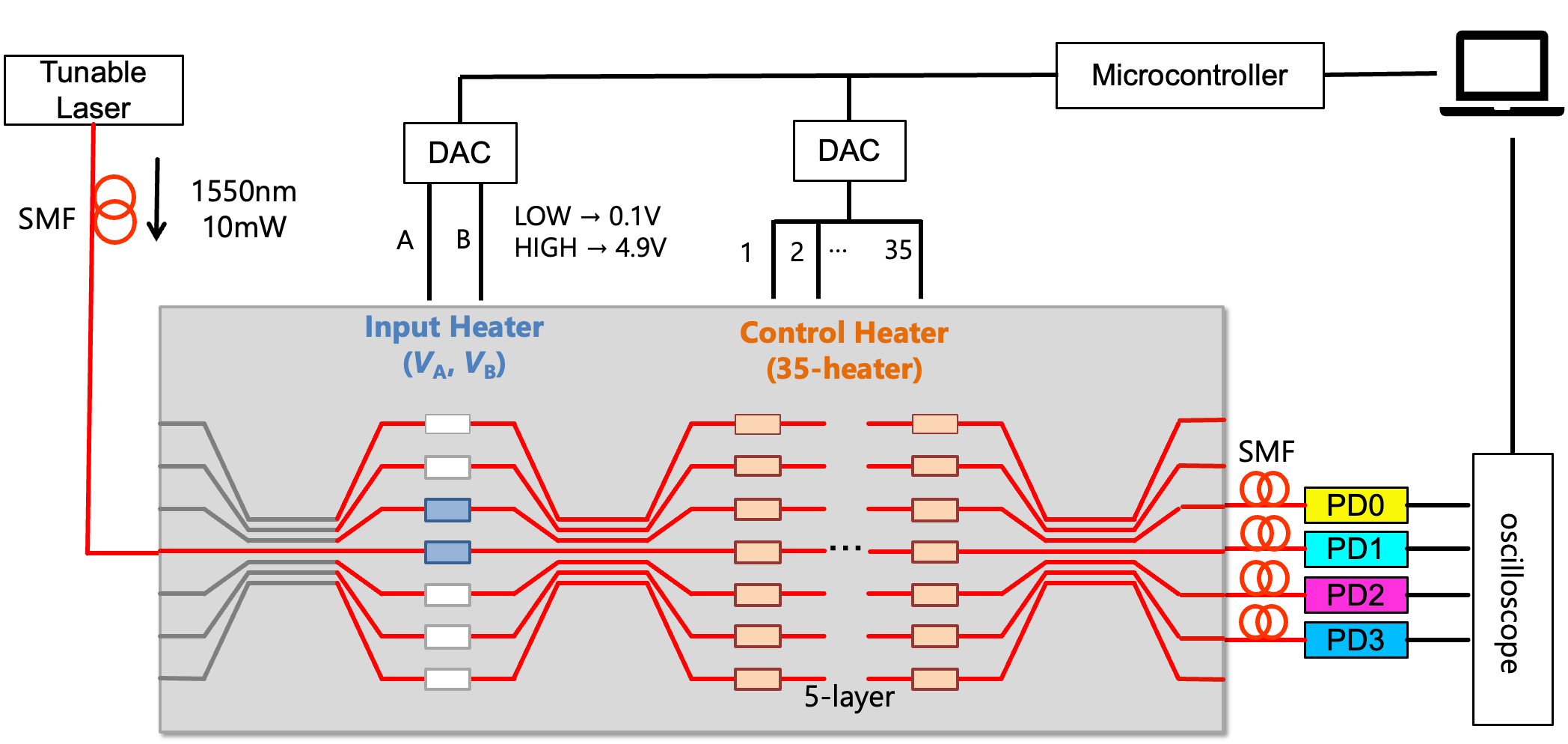}
    \caption{Experimental setup used for optical logic gate and decoder measurements with closed-loop control.}
    \label{fig:experimental_setup}
\end{figure}

The experimental setup is shown in Fig.~\ref{fig:experimental_setup}. Light from a tunable laser source (1550\,nm, 10\,mW) is coupled into the central input port. Logic signals are encoded via phase modulation rather than dedicated physical input ports. Specifically, the two heaters in the middle channels of the third coupler stage define inputs A and B: voltages of 0.1\,V and 4.9\,V correspond to logic ``0'' and ``1,'' respectively. These values were chosen arbitrarily and are parameters for the optimization; therefore, they can be modified.

To determine the heater settings required for a given logic function, we employ Bayesian optimization. Candidate voltage configurations are generated on a computer and transmitted to a microcontroller, which drives the D/A board. The optical outputs are monitored in real time by InGaAs photodetectors, digitized by an oscilloscope, and sent back to the computer. This closed-loop feedback enables continuous optimization of the phase settings, ensuring reliable performance despite fabrication imperfections and environmental variations.

\section{Results}
\subsection{Optical Logic Gates}

Six reconfigurable optical logic gates were experimentally demonstrated using the programmable architecture and the optimization method described above. For each target function, the algorithm identified suitable voltage configurations that maximized extinction ratios and ensured consistent switching behavior. Only one output port was used for the optical logic gates, the optimization can be adjusted for any of the ports.

\begin{figure}[h!]
    \centering
    \includegraphics[width=1\columnwidth]{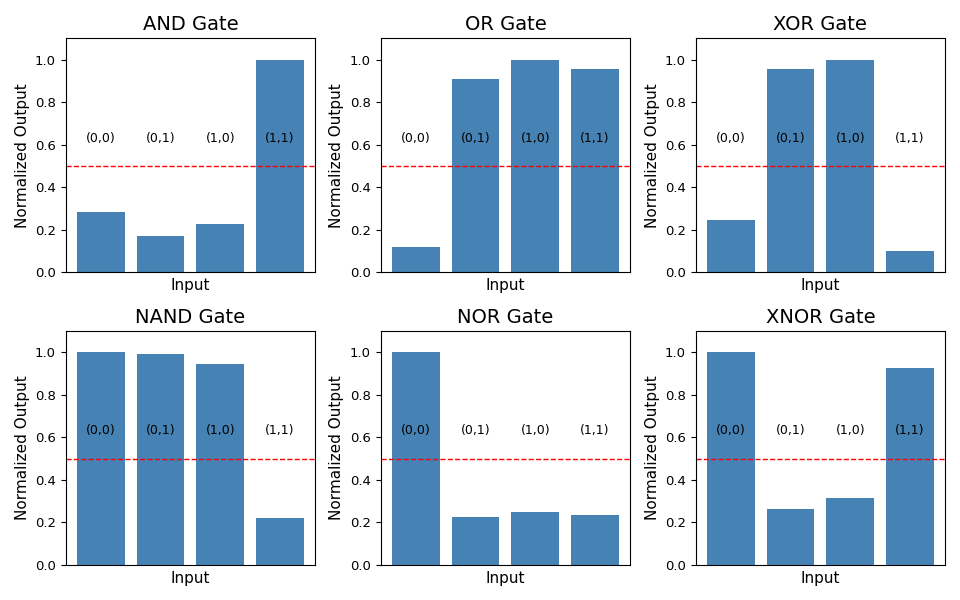}
    \caption{Measured output intensities for six optical logic gates under optimized heaters configuration.}
    \label{fig:logic_gates}
\end{figure}

\begin{figure}[h!]
    \centering
    \includegraphics[width=1\columnwidth]{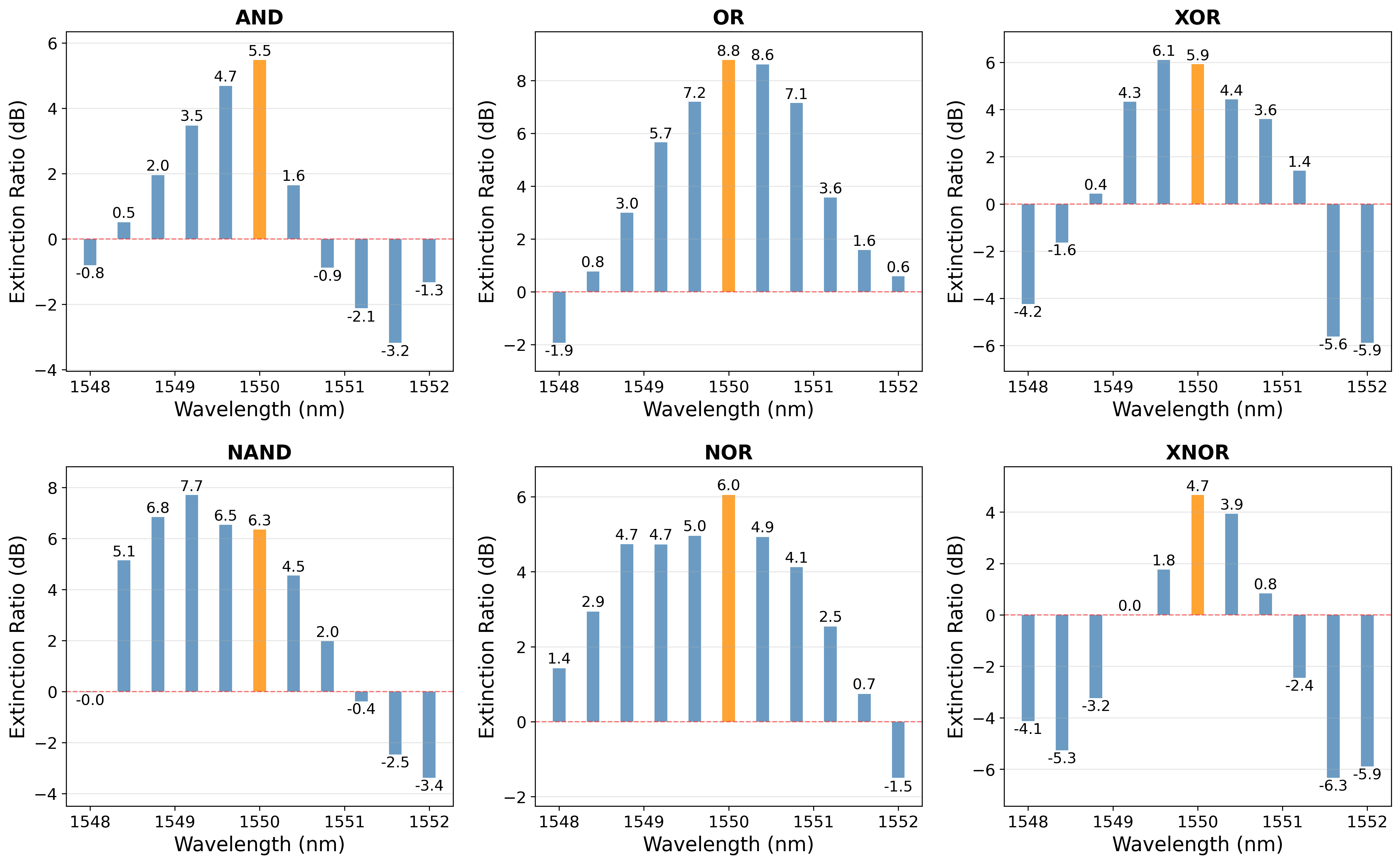}
    \caption{Extinction ratio of the optical logic gates measured across multiple wavelengths with 50~GHz spacing.}
    \label{fig:ER_Gates}
\end{figure}

Figure~\ref{fig:logic_gates} presents the measured output intensities for the implemented gates under optimized conditions. In all cases, the outputs corresponding to logical ``0'' and ``1'' are clearly separated, confirming correct gate operation. To further characterize the device, we evaluated the same phase configurations across multiple wavelengths with 50~GHz channel spacing. The results, summarized in Fig.~\ref{fig:ER_Gates}, show that while extinction ratios vary with wavelength, the logic functionality is preserved over a limited spectral range. This wavelength-dependent behavior indicates the potential for parallel or wavelength-multiplexed operation, although achieving uniform performance across the full spectrum would require additional optimization.

\subsection{2-Bit Decoder}

To further demonstrate the versatility of the platform, we implemented a 2-bit optical decoder on the same device using the same optimization method. The two input bits were encoded through thermo-optic phase shifters, four output ports were used, the optimization can be adjusted according to the ports selected.

Bayesian optimization identified heater voltage configurations that produced clear one-hot output patterns. Table~\ref{tab:Decoder_resuts} summarizes the measured output intensities for all four input states. In each case, a single output channel dominates, confirming correct decoding behavior.

\begin{figure}[h!]
    \centering 
    \includegraphics[width=0.95\columnwidth]{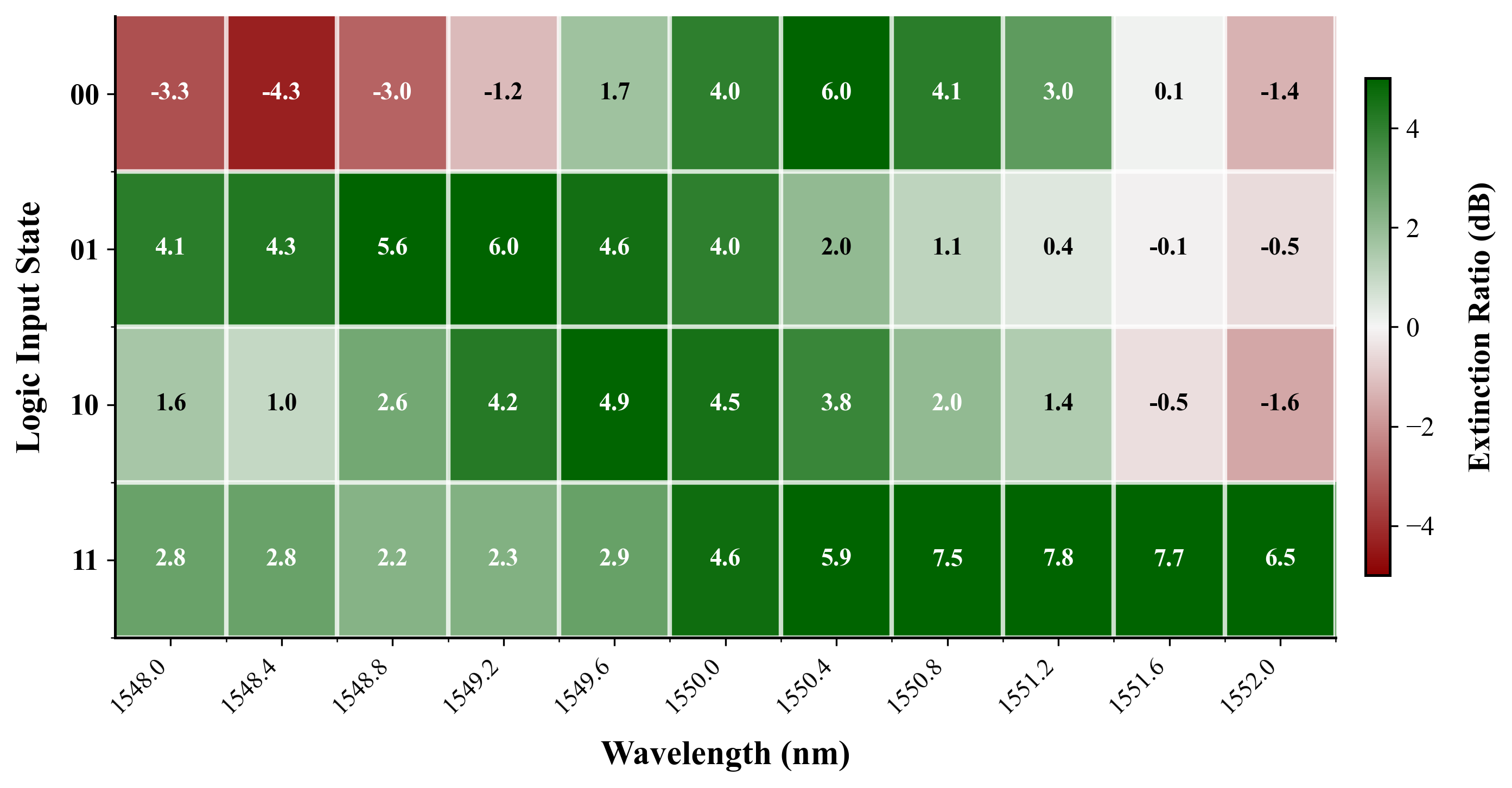} 
    \caption{2-bit Decoder performance across wavelengths with 50~GHz spacing.}
    \label{fig:decoder_acrossWV}
\end{figure}

The decoder was also characterized across multiple wavelengths with 50~GHz spacing. As shown in Fig.~\ref{fig:decoder_acrossWV}, the correct output channels retained their relative dominance, even though extinction ratios varied with wavelength. This wavelength-dependent response indicates that the reconfigurable decoder can operate over a limited spectral window, supporting scalability toward multi-channel or wavelength-parallel processing.

\begin{table}
\caption{2-bit Decoder Performance}
\label{tab:Decoder_resuts} 

\begin{tabular}{|c|c|c|c|c|c|}
\hline
A & B & $V_A(V)$ & $V_B(V)$ & Output Voltage (V) & ER [dB] \\
\hline
\adjustbox{valign=c}{0} & \adjustbox{valign=c}{0} & \adjustbox{valign=c}{0.1} & \adjustbox{valign=c}{0.1} & \adjustbox{valign=c}{\includegraphics[width=1in]{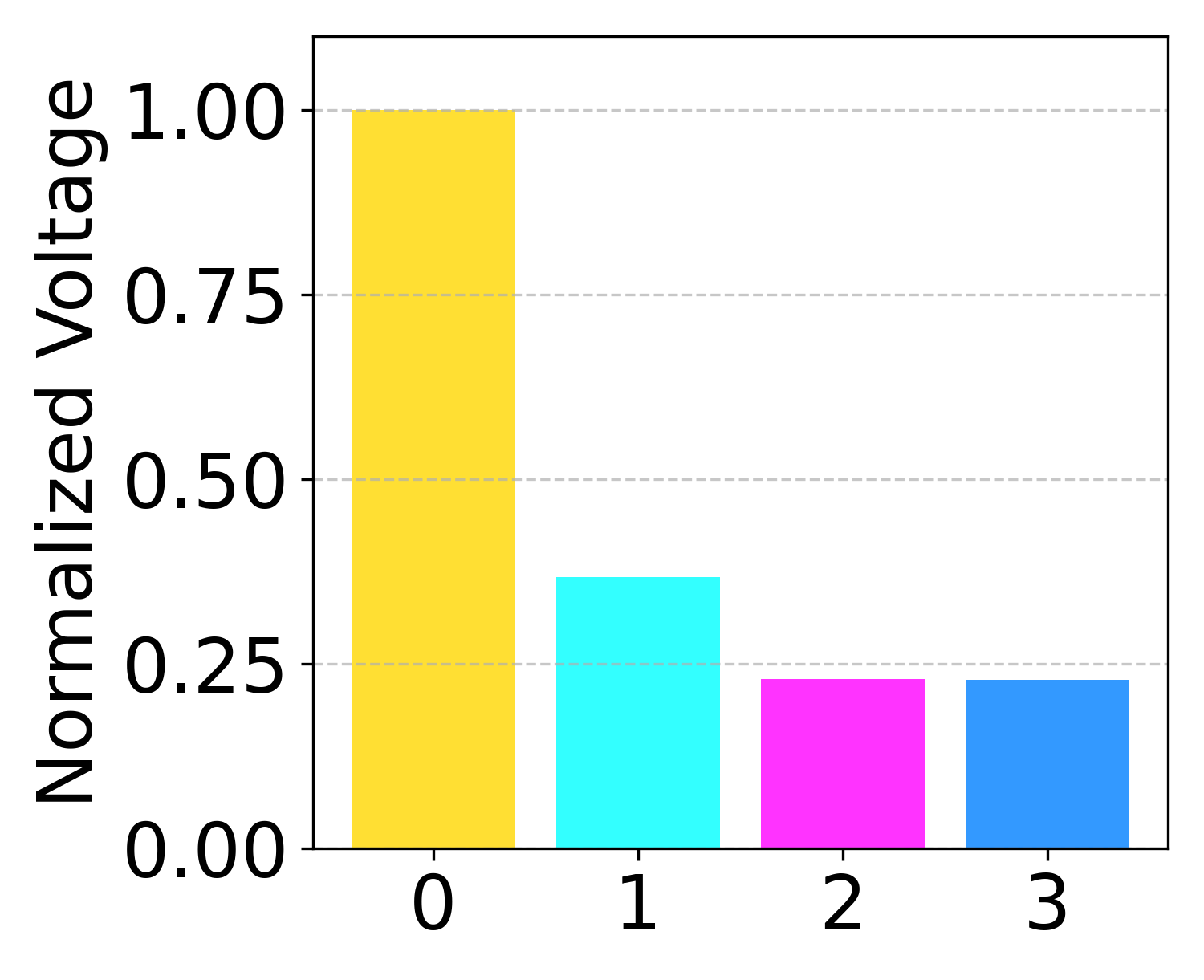}} & \adjustbox{valign=c}{4.35} \\
\adjustbox{valign=c}{0} & \adjustbox{valign=c}{1} & \adjustbox{valign=c}{0.1} & \adjustbox{valign=c}{4.9} & \adjustbox{valign=c}{\includegraphics[width=1in]{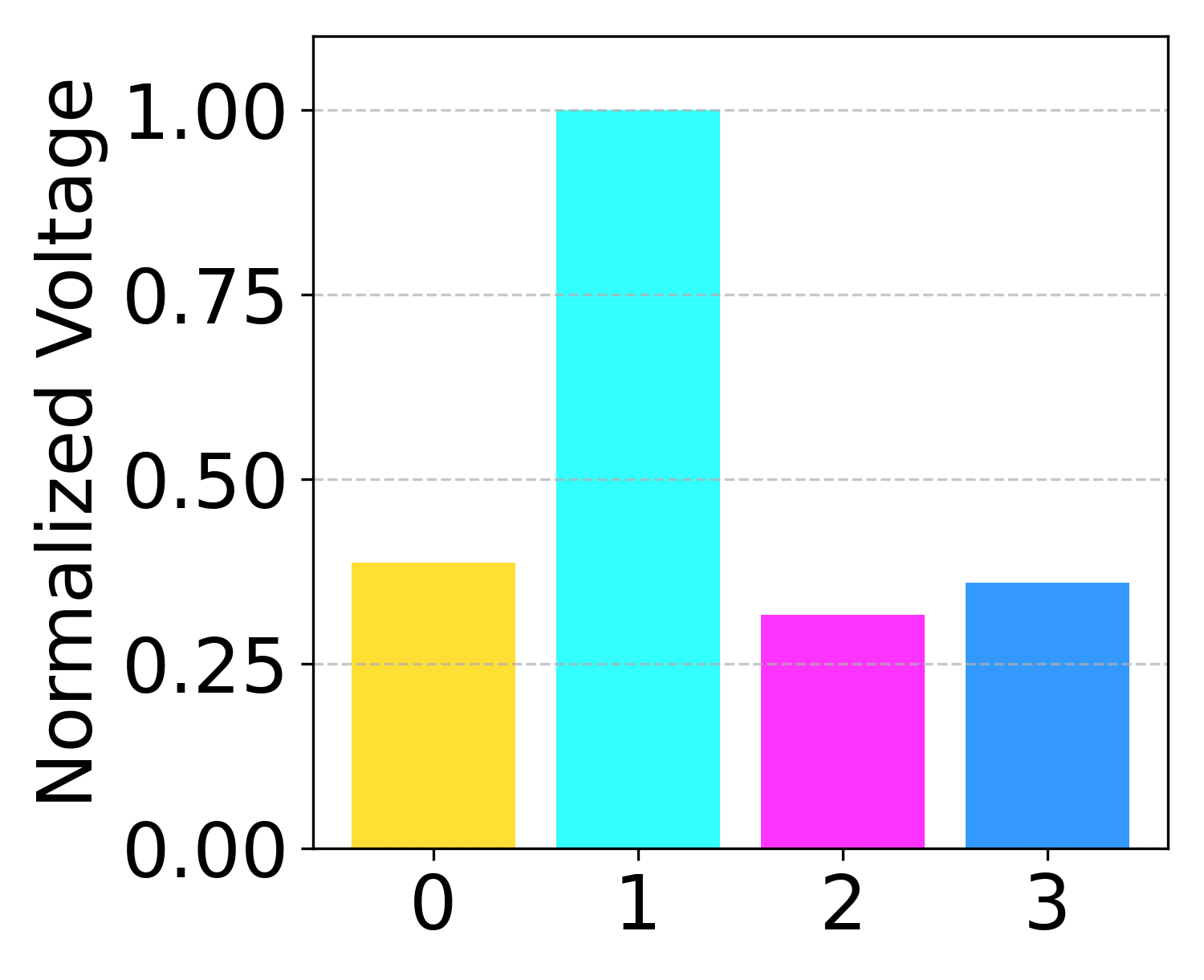}} & \adjustbox{valign=c}{4.13} \\
\adjustbox{valign=c}{1} & \adjustbox{valign=c}{0} & \adjustbox{valign=c}{4.9} & \adjustbox{valign=c}{0.1} & \adjustbox{valign=c}{\includegraphics[width=1in]{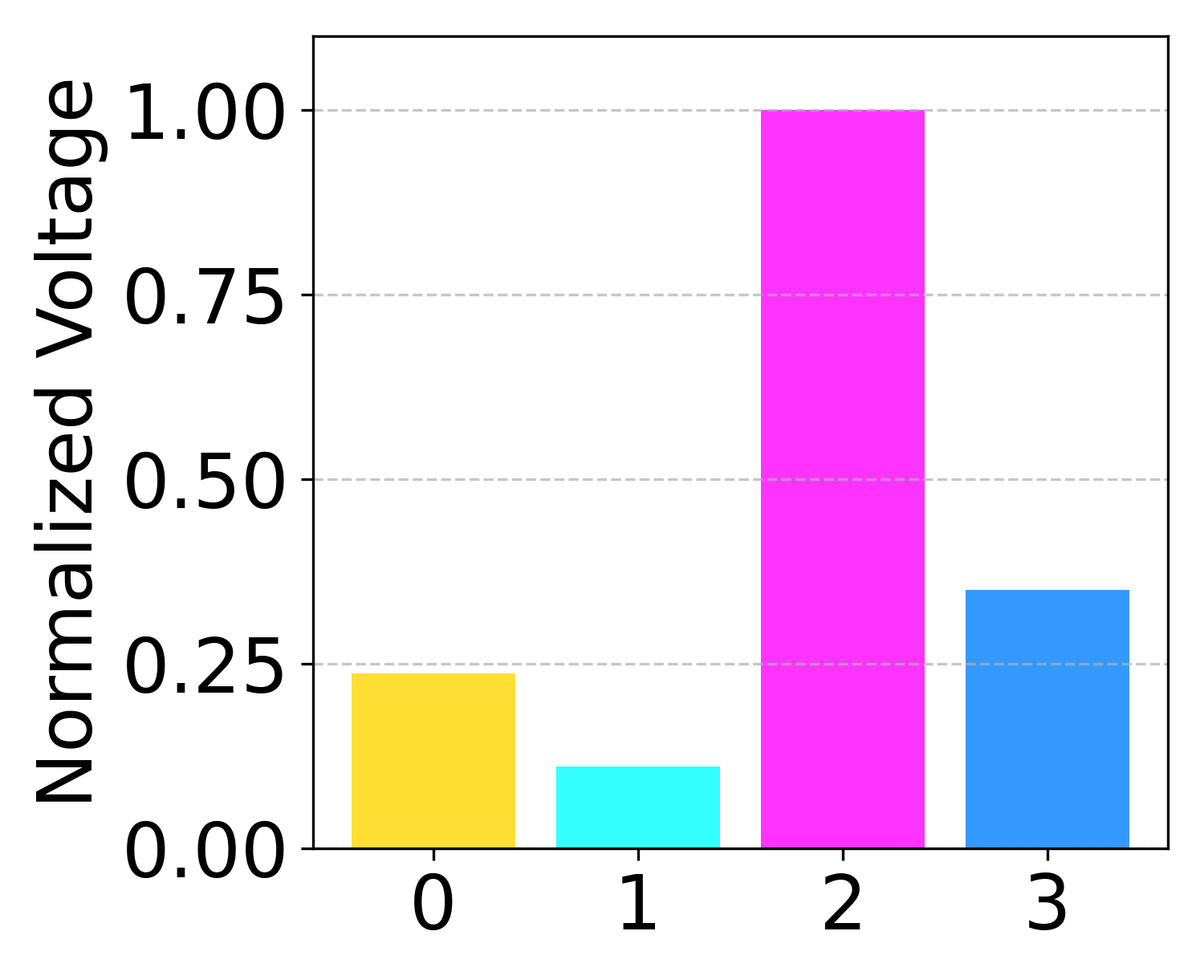}} & \adjustbox{valign=c}{4.41} \\
\adjustbox{valign=c}{1} & \adjustbox{valign=c}{1} & \adjustbox{valign=c}{4.9} & \adjustbox{valign=c}{4.9} & \adjustbox{valign=c}{\includegraphics[width=1in]{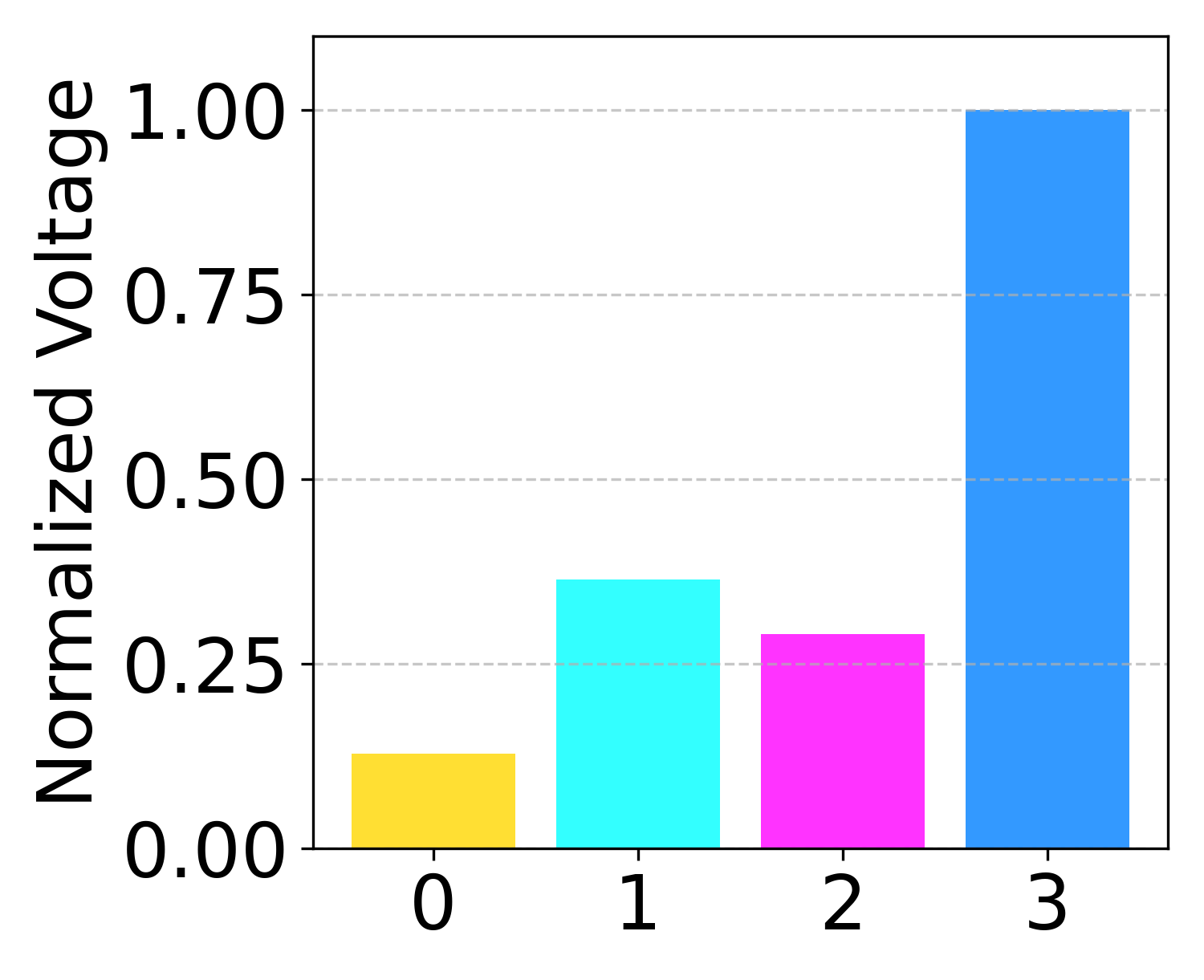}} & \adjustbox{valign=c}{4.38} \\
\hline
\end{tabular}

\end{table}

\section{Conclusions}

We have presented a reconfigurable photonic platform that implements both optical logic gates and a 2-bit decoder using cascaded multiport directional couplers and thermo-optic phase shifters. 

In contrast to prior fixed-function optical circuits, the proposed approach enables multiple logic functions and decoding operations on the same hardware through reconfiguration. Operation across several wavelengths further indicates the potential for wavelength-multiplexed optical computing. 

Future work will focus on the implementation of different logic components, and faster tuning mechanisms. Overall, these results highlight a pathway toward flexible photonic processors for next-generation information systems.

\bibliography{References/references}
\bibliographystyle{IEEEtran}

\section*{Declaration of Interests}
The authors declare that they have no known competing financial interests or personal relationships that could have appeared to influence the work reported in this paper.

\vfill

\end{document}